# Artificial Intelligence in Financial Forecasting: Analyzing the Suitability of AI Models for Dollar/TL Exchange Rate Predictions

*Asef Yelghi, Aref Yelghi, Shirmohammad Tavangari*

1. Institute of *banking* and Insurance, Department of Banking, Marmara University, Istanbul,Turkey.
asefyelghi@marun.edu.tr

2. Deprtement of Computer Engineering, Istanbul Topkapi University Istanbul, Turkey
aref.yelghi@topkapi.edu.tr

3. Deprtement Electrical and Computer Engineering, University of British Columbia Vancouver,BC,Canada
s.tavangari@alumni.ubc.ca

**Abstract**

The development of artificial intelligence has made significant contributions to the financial sector. One of the main interests of investors is price predictions. Technical and fundamental analyses, as well as econometric analyses, are conducted for price predictions; recently, the use of AI-based methods has become more prevalent.

This study examines daily Dollar/TL exchange rates from January 1, 2020, to October 4, 2024. It has been observed that among artificial intelligence models, random forest, support vector machines, k-nearest neighbors, decision trees, and gradient boosting models were not suitable; however, multilayer perceptron and linear regression models showed appropriate suitability and despite the sharp increase in Dollar/TL rates in Turkey as of 2019, the suitability of valid models has been maintained.

**Introduction**

In recent years, the development of machine learning has gained momentum across various sectors, including financial markets. Fintech companies are contributing to the efficient and effective functioning of the financial system. One of the primary interests of investors is price prediction. Various approaches, models, and applications have been developed for price forecasting. However, artificial intelligence models have introduced different paradigms, focusing on the development of appropriate models by addressing functions similar to human intelligence.

The analysis of financial markets has begun to be interpreted using econometric analyses. With the advancement of technology, algorithms have started to play a role in these models, leading researchers to adopt new methodologies. Additionally, a comparative analysis of artificial intelligence models will be conducted. This study will perform a comparative analysis of machine learning models, focusing on the Dollar/TL exchange rate. Forecasting financial markets is regarded as a significant issue, prompting the development of new approaches and models to address this problem.

The research examines daily Dollar prices from January 2020 to October 4, 2024, aiming to identify price behaviors using artificial neural network models. Furthermore, the suitability of various artificial intelligence models for making predictions will be evaluated.

**A. Practical Studies on Machine Learning**

The literature review reveals that many studies have begun to utilize artificial intelligence models in financial markets. In their study, Donaldson and Kamstra (1996) applied an artificial neural network model to forecast short-term interest rates in Canada using a dataset from the 1970s to 1980s. The findings demonstrated that artificial neural networks provided more accurate results in macroeconomic forecasting compared to traditional methods.

Patel et al. (2014) examined the stock prices of Reliance Industries and Infosys Ltd., as well as the CNX Nifty 30 and S&P Bombay Stock Exchange (BSE) Sensex indices, between 2003 and 2012. This study utilized Artificial Neural Networks (ANN), Support Vector Machines (SVM), Random Forest, and Naive-Bayes models for predictions. The results indicated that the Random Forest model outperformed the other three prediction models in overall performance.

In the study by Ding et al. (2015), financial market predictions were made using news articles collected from agencies such as Reuters and Bloomberg. This research employed convolutional neural networks (CNN) and deep learning models, analyzing data from 2005 to 2014. The results showed that CNN and deep learning models performed better compared to traditional market forecasting methods.

Lai et al. (2018) utilized financial time series data (e.g., stock prices) alongside macroeconomic indicators in their study. The dataset is generally sourced from financial data providers such as Yahoo Finance or Bloomberg. The relationships from the 2000s to the 2010s were successfully captured by Deep Neural Networks (DNN) and LSTM methods, leading to improved prediction accuracy.

In the study by Lee and Yoo (2019), Deep Neural Networks (DNN) were applied to the KOSPI, S&P 500, DAX, and NASDAQ indices. This research, conducted with data collected between January 1, 2006, and December 31, 2017, demonstrated that jointly assessing international stock markets could enhance prediction accuracy, showing that deep neural networks are highly effective for such tasks.

Yunus EmreGür (2024) used Support Vector Machines (SVM), Extreme Gradient Boosting (XGBoost), Long Short-Term Memory (LSTM), and Gated Recurrent Units (GRU) methods for predicting the Euro. This study, which analyzed data from January 2000 to October 2023, found that GRU outperformed other measurements in terms of accuracy.

Yelghi and other researchers (2024) introduced new paradigms focusing on the development of meta-heuristic algorithms and their integration with financial modeling. In a previous study (Yelghi et al., 2021), they examined the relationship between interest rates, determined by loan types in the banking sector, with inflation and exchange rates, providing significant findings in this area. In another study (Yelghi et al., 2022), they aimed to improve the performance of ANFIS models through the integration of meta-heuristic algorithms, achieving highly successful prediction results.

In their research on forecasting operations, Yelghi and Yelghi (2024) proposed the Wave Net-TSRS model, offering an innovative approach to predicting financial time series and obtaining significant results with this deep learning method. Furthermore, Yelghi et al. (2024) developed an algorithm based on a happiness model that can be applied in prediction processes.

Tavangari et al. (2022) highlighted the characteristics of meta-heuristic algorithms and developed a new integration model. Asherlou et al. (2022) used meta-heuristic algorithms to develop a new optimal path alignment approach in the construction sector. Erkan et al. (2024) created a new prediction model for algorithmic trading using the hill climbing algorithm.

**Table 2. Literature Review**

| Author | Year | Variable | Model | Dataset | Findings |
|---|---|---|---|---|---|
| Donaldson & Kamstra | 1996 | Forecasting Canada's short-term interest rates | Artificial Neural Networks | 1970s and 1980s | Neural networks provided more accurate results in macroeconomic forecasting. |
| Patel et al. | 2014 | Two stocks, Reliance Industries and Infosys Ltd., and two indices, CNX Nifty 30 and S&P BSE Sensex | ANN, SVM, Random Forest, Naive-Bayes | 2003-2012 | The Random Forest model outperformed the other three prediction models in overall performance. |
| Ding et al. | 2015 | News texts collected from agencies like Reuters and Bloomberg | Convolutional Neural Networks (CNN), Deep Learning Models | 2005-2014 | CNN and deep learning models exhibited better performance compared to traditional market forecasting methods. |
| Chong et al. | 2017 | Korean Stock Exchange | DNN, ANN | 2013-2014 | Deep neural networks were able to extract additional information from the residuals of autoregressive models, improving prediction performance. |
| Heaton et al. | 2017 | Historical price data from U.S. stock markets | Deep Learning Models for Portfolio Optimization | 1990s to 2010s | Suggested that these models could serve as alternatives to classical financial theories. |
| Qin et al. | 2017 | Data from the Chinese stock market | LSTM, Attention Mechanism | 2000-2015 | The attention mechanism effectively highlighted critical time points. |
| Fischer & Krauss | 2018 | Historical daily closing prices of stocks in the | Long Short-Term Memory (LSTM) | 1992-2015 | Achieved better results in processing complex time |

| Author | Year | Variable | Model | Dataset | Findings |
|---|---|---|---|---|---|
| | | S&P 500 index | | | series data. |
| Namini | 2018 | Datasets including financial market indices, exchange rates, and macroeconomic indicators | ARIMA, LSTM | 2000s to 2010s | LSTM demonstrated superior performance in long-term forecasting. |
| Pendharkar | 2018 | Data obtained from Bloomberg, CRSP, or other financial data providers | Reinforcement Learning | 1990s to 2010s | Highlighted adaptability to changing market conditions. |
| Moody &Saffell | 2018 | Data including exchange rates, stock indices, and bond yields | Feedback Neural Networks, Reinforcement Learning | 1980s and 1990s | The direct reinforcement model successfully improved profit performance in trading decisions. |
| Lai et al. | 2018 | Financial time series (e.g., stock prices) and additional macroeconomic indicators | Deep Neural Networks, LSTM | 2000s and 2010s | Improved prediction accuracy by successfully capturing patterns across different time scales. |
| Lee &Yoo | 2019 | KOSPI, S&P 500, DAX, and NASDAQ | DNN | January 1, 2006 - December 31, 2017 | Joint assessment of international stock markets can enhance prediction accuracy; deep neural networks are highly effective for such tasks. |
| Yunus EmreGür | 2024 | Euro | SVM, XGBoost, LSTM, GRU | January 2000 - October 2023 | Found that the GRU algorithm provided better accuracy than the others. |

As observed in the literature, financial markets, including stock exchanges and currency markets, have been studied from various perspectives. Generally, the use of artificial neural networks in financial markets has yielded more favorable results. This study aims to investigate which artificial intelligence model is most suitable through their application.

**B. Application of Artificial Intelligence Models**

In this study, various artificial intelligence models, including Random Forest, Support Vector Machines, Multi-Layer Perceptron, K-Nearest Neighbors, Decision Tree, Gradient Boosting, and Linear Regression, will be applied. Initially, the data loading process was performed. In this step, relevant date and price information for the U.S. dollar against the Turkish lira was loaded from the appropriate sheet of an Excel file using the pandas library.

Subsequently, during the data preparation phase, the date information was converted to datetime format and set as the index. In this study, the date was used as an index because it plays a crucial role in determining the sequence during the analysis process. Next, during the feature engineering stage, lag features (lag1 and lag2) were created, and a 7-day moving average (rolling_avg) was calculated. These features help the model make more accurate predictions by leveraging past price movements. The results obtained from the application of artificial intelligence models are illustrated in the following Graph 1.

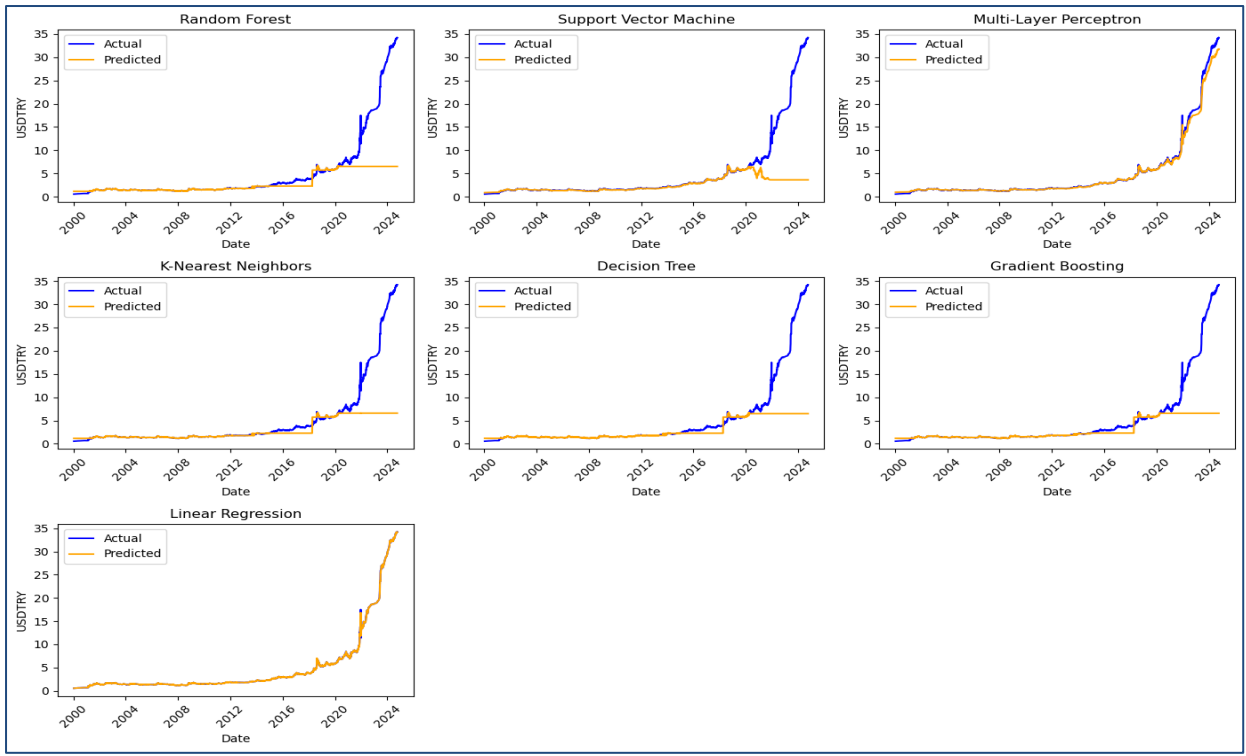

**Figure 1.**

Looking at the graphs above, it can be observed that there is a significant divergence between the actual values and the predicted values for the Random Forest, Support Vector Machines, K-Nearest Neighbors, Decision Tree, and Gradient Boosting models, excluding the Multi-Layer Perceptron and Linear Regression models, during the years 2018-2019. The acceleration of the dollar's rise during this period indicates that these four models were not suitable for predicting the dollar's excessive valuations. However, it has been found that the Multi-Layer Perceptron and Linear Regression models maintained their suitability despite the acceleration in the dollar's rise.

On the other hand, the statistical results from the models confirm this finding. The model's suitability is ideally indicated by a high $R^2$ value, demonstrating strong explanatory power, while lower values for risk metrics such as RMSE and MAE are expected. The statistical findings are presented in the table below.

## Table 2. Results of the Models

| Machine Learning Algorithms | RMSE | MAE | R² |
|---|---|---|---|
| Random Forest | 3.006157 | 2.210849 | 0.335757 |
| Support Vector Machines | 3.25585 | 2.517527 | 0.35338 |
| Multi-Layer Perceptron | 0.322154 | 0.241591 | 0.926885 |
| K-Nearest Neighbors | 2.9964 | 2.19809 | 0.335199 |
| Decision Tree | 3.014639 | 2.220644 | 0.323587 |
| Gradient Boosting | 2.99365 | 2.197563 | 0.343187 |
| Linear Regression | 0.056528 | 2.197563 | 0.99601 |

According to the table above, the Multi-Layer Perceptron and Linear Regression models exhibit higher explanatory power and lower risk metrics, as indicated by RMSE and MAE. In contrast, the Random Forest, Support Vector Machines, K-Nearest Neighbors, Decision Tree, and Gradient Boosting models show both lower explanatory power and higher risk metrics.

**Conclusion and Evaluation**

Price forecasting is a primary area of interest for investors in the market. To address this, numerous studies and methodologies have been developed, and analyses have been conducted. Recently, the advancement of artificial intelligence has led to an increased focus on price prediction. According to the literature, AI methods have been shown to yield more favorable results compared to econometric analysis findings.

This study presents a comparative analysis of the models used for predictions in artificial intelligence. Daily data for the Dollar/TL exchange rate from the year 2000 onward has been utilized. The findings indicate that the models—including Random Forest, Support Vector Machines, K-Nearest Neighbors, Decision Tree, Gradient Boosting, and Linear Regression—yielded predictions that closely aligned with actual values up to 2019, as depicted in the graphs. However, a significant divergence between actual values and predicted values emerged starting in 2019. Only the Multi-Layer Perceptron and Linear Regression models demonstrated successful and meaningful results in price predictions, despite the dollar's excessive appreciation.